\documentclass[prl,twocolumn,showpacs]{revtex4-1}
\usepackage{array,amsmath,amssymb,graphicx}

\newcommand{\LL}{{\cal L}}
\newcommand{\EE}{{\cal E}}

\usepackage{array,color,amsmath,amssymb,graphicx}

\begin{document}
\title{Phase Diagram of Electron Systems near the Superconductor-Insulator
Transition}
\author{V.L. Pokrovsky $^{1,2}$}\author{G.M. Falco$^3$}
\author{T. Nattermann$^3$}\affiliation{$^1$Department of Physics, Texas A\&M University, College Station, Texas
77843-4242}
\affiliation{$^2$Landau Institute for Theoretical Physics, Chernogolovka, Moscow District,
142432, Russia}
\affiliation{$^3$Institut f\"ur Theoretische Physik, Universit\"at zu K\"oln, Z\"ulpicher
Str. 77, D-50937 K\"oln, Germany}
\date{\today}

\begin{abstract}
The zero temperature phase diagram of Cooper pairs exposed to disorder and
magnetic field is determined   theoretically from a variational approach.  Four distinct phases are found: a Bose and a Fermi
insulating, a metallic and a superconducting phase, respectively. The
results explain the giant negative magneto-resistance found experimentally
in In-O, TiN, Bi and high-$T_c$ materials.

\end{abstract}

\maketitle

\affiliation{Department of Physics, Texas A\&M University, College Station, Texas
77843-4242}
\affiliation{Landau Institute for Theoretical Physics, Chernogolovka, Moscow District,
142432, Russia}
\affiliation{Institut f\"ur Theoretische Physik, Universit\"at zu K\"oln, Z\"ulpicher
Str. 77, D-50937 K\"oln, Germany}

\looseness=-3\noindent \emph{Introduction.} Some alloys such as InO$_{x}$ \cite%
{shahar-92,hebard,gant-2000,gant-01}, TiN \cite{baturina}, BiSr$_{2}$Ca$_{z}$%
Pr$_{1-z}$Cu$_{2}$O$_{8+y}$ \cite{beschoten}\ as well as ultrathin films of
Bi \cite{goldman}, Be \cite{adams,butko,wu} and of high-T$_{c}$ superconductors
(SC) \cite{beschoten,valles}\ display (quantum) superconductor to insulator phase
transition (SIT)
(for a recent
review see\cite{review}).
Besides of the very existence of the SIT, the
experiments demonstrate several unusual phenomena in its vicinity. The most
general is the
maximum of the resistivity in external
magnetic field in the insulator phase: at growing field the resistance first
grows, sometimes by several orders of magnitude, and then strongly decreases
\cite{wu,Gantmakher-96,gant-2000,shahar-04,valles-09}.
This surprising decrease
is called giant negative magnetoresistance (GNM). In  \cite%
{Gantmakher-96, adams-pairs, valles-09} the GNM was considered as %
a signature of localized Cooper pairs (CP) %
surviving in the insulating state and their decay in strong enough magnetic field.
In the most thoroughly
studied amorphous alloy InO$_{x}$ experimenters observed a transition from
an insulating to a metallic phase at increasing magnetic field \cite%
{gant-2000}, and from superconductor to metal if magnetic field and density
of carriers grow simultaneously \cite{kapitulnik}. In the same films of InO$%
_{x}$, in the insulating state at large magnetic field of about 15T the
observed resistivity obeyed Mott's $3d$ variable hopping law $R\propto \exp
(T_{0}/T)^{1/4 }
$ \cite{Gantmakher-96} which
shows that the Coulomb interaction was screened, and that InO$%
_{x}$ films of thickness $w\approx 20nm$ used in \cite{Gantmakher-96} must
be considered as a 3d object. This implies that the essential length scales
in this alloy were less or of the order of 20$nm$.

\looseness=-20 There are several theoretical approaches to the SIT. One of them
is based on the BCS theory and accounts for the Coulomb interaction
enhanced by disorder \cite{finkelstein}. It shows that in $d=2$ even
weak Coulomb interaction may turn the transition temperature to zero and
that appearing fermions are localized. Another idea specific for
granulated superconductors \cite{efetov} suggests that the CP are bound in granules and
can tunnel between them. Depending on relative strength of Coulomb
interaction and tunneling amplitude, either the insulating or the
superconducting state is realized. In \cite{Shklovskii-09} the SIT line as a
function of the CP density $n_{b}$ was calculated for a system of charged
fermions bound into pairs of a
fixed size in the field of randomly distributed Coulomb centers. In \cite{fisher}
the idea of duality between vortices and CP was proposed and
some exact results where found. Neither of these theories explains the GNM.
Numerical calculations
of the Bogoliubov-de Gennes equations in a random medium \cite{ghosal}
showed the appearance of superconducting islands separated by the insulating
see. This islands were shown to be gradually suppressed by magnetic field
\cite{meir2}.

\looseness=-20\noindent In this article we propose a theory of insulating state near the
SIT in magnetic field. We assume that the CP
survive in the insulating state, but are localized by a random Gaussian potential. The model also incorporates the CP interaction and the
magnetic field which destroys the pairs when strong enough. Our theory
explains the GNM and the metal insulator transition (MIT) in growing magnetic field and predicts new phases:
the Bose insulator (BI) and the Fermi insulator (FI) \cite{priority}.
\looseness=-3 Our physical picture is that in systems displaying the SIT the Fermi energy $ E_F$
is close to the mobility threshold such that $k_Fl\approx 1$. Here $l$ is mean free path in the normal
state near the SIT.
Therefore, the electron density
fluctuations on large scales like coherence length $\xi $ and Larkin length
$\mathcal{L}_b  (>\xi>l)$  for CP (see below) are relatively small. This conclusion is
supported by very short electron density correlation length found in \cite%
{ghosal}. The disorder does not influence substantially neither the number
of CP
nor their binding energy  $\Delta $.
However,
the density of CP $n_b\sim n_e\Delta /E_F$ is much smaller than the density ${n_e}$ of the background
electrons. Therefore the random potential  can substantially change the positions
of centers of mass of CP and localize them. The Coulomb interaction at low $T$
seen in experiments is screened on distances exceeding the
electron distance  $n_{e}^{-1/d}\approx \pi/k_F$, which is
about $1nm$ \cite{shahar-92,gant-2000,gant-01}).
Thus these length scales are less than the CP spacing
and the film thickness $w$.
\looseness=-20The Gaussian random short-range potential has a characteristic localization length $%
\mathcal{L}$ (Larkin length) \cite{Falco2}. The random
field acting on interacting particles is renormalized by the self-consistent field
and changes with energy. The CP and Fermi-excitations appearing at their decay
are separated by energy gap from the background electrons. Therefore, the random
field acting on the CP and excitations is
weaker than the field acting on the background electrons.
Deeply
localized pairs are confined in fluctuation potential wells whose size is
smaller than $\mathcal{L}_b$. In such a small well the CP interaction is
essential.

\noindent
\emph{The model and general approach.}
\looseness-20 Let a system of charged particles (electrons or CP) in a Gaussian random potential $U(\mathbf{r})$
to be exposed to a homogeneous magnetic field $\mathbf{B}$. The single-particle Hamiltonian is given by
\begin{equation}
\hat{\mathcal{H}}=\frac{\hbar ^{2}}{2m_{k}}\left[-\mathbf{\nabla }^{2}+\left(
\frac{e_{k}}{2\hbar c}\right) ^{2}({\mathbf{r}}\times \mathbf{B})^{2}\right]+U_{k}(%
\mathbf{r}),
\end{equation}%
\looseness-20 where $k=b,f$ for bosons and fermions, respectively, and $m_{f}=m_{b}/2=m$
and $e_{f}=e_{b}/2=e$. The 
spin of bosons is zero. The random potential has zero average and the
pair correlator
$\langle U_{k}(\mathbf{r})U_{k}(\mathbf{r}%
^{\prime })\rangle ={\kappa _{k}^{2}}\delta (\mathbf{r}-\mathbf{r}^{\prime
}).$
The stray random potential for electrons
is roughly twice smaller than that for bosons,
 i.e. $\kappa _{b}\approx 2\kappa _{f}$. %
 Two relevant length scales are the magnetic length
 $\ell _{k}=\sqrt{\hbar c/(e_{k}B)}$
and the Larkin length
$\mathcal{L}_{k}={ \sqrt{\pi}(\sqrt{2}\pi)^{d-2}}(\hbar ^{2}/(m_{k}\kappa _{k}))^{2/(4-d)},$
  where %
the electron Larkin length
$\mathcal{L}_{f}$ is $2^{4/(4-d)}$ times larger than
the CP length $\mathcal{L}_{b}$.  At low magnetic fields and 
 strong disorder, the CP fill rare deep wells of
the random potential. We call this state the {\it Bose insulator}. When the disorder decreases to
a critical value, the potential wells overlap and transition to superconducting
phase proceeds.  In the insulating phase a strong magnetic field destroys the CP by one of two mechanisms: (i)
Paramagnetic depairing at which Zeeman energy of two electrons exceeds the
CP binding energy renormalized by disorder
and (ii) Diamagnetic squeezing at
which the size $R$ of the optimal potential wells becomes less than the size $\xi$ of the CP. In a dirty superconductor $\xi=0.85\sqrt{3\xi_0 l/d}$, where $\xi_0=\hslash v_F/(\pi\Delta)$ is the BCS coherence length and $l$ is the electron mean free path.
Which effect, (i) or (ii),  dominates depends
on $d$, strength of disorder and the electron density. At higher magnetic fields a fraction of CPs decays
into fermions.
As long as the density of excited fermions remains small,
they are also localized. This state we call {\it Fermi insulator}.  The paramagnetic depairing  transition
from Bose to Fermi insulator (BFT) happens when the energy of the CP becomes equal to
energy of appearing two electrons:
\vspace{0cm}
\begin{equation}\quad { \mu_{b}}-{2}\Delta =2\left( E_{f}+E_{z}\right) ,  \label{BFT}
\vspace{0cm}
\end{equation}
\looseness-5 where ${ \mu}_{b}$ is {the chemical potential  of the CP at fixed density} $n_b$
and magnetic field $\mathbf{B}$, and $2\Delta $ is {their} binding energy; $E_{f}$
is the disorder energy for an appearing fermion and $E_{z}=-{g_e}\mu _{B}B/2$
is its Zeeman energy.
At further increase
of magnetic field the density of fermions reaches  $n_{fc}={\cal L}_f^{-d}$%
. Then their wave functions strongly overlap and their interaction
is strong enough to overcome the localization provided
$\mathcal{L}_f$ exceeds the Bohr's radius in the media.
Delocalization of
fermions turns the FI into a metal, this is the MIT.

\noindent                             \emph{Thin films in parallel field.}
\looseness=-30 We first analyze the BFT
equation for the case of a very thin film of thickness $w (\ll \ell ,%
\mathcal{L})$ in a  parallel magnetic field. In this case the
field does not produce any diamagnetic effect, only paramagnetic one.
The energy of the first appearing fermion $E_{f}$ can be found from the
condition that a boson trapped by the same potential well as the fermion has
energy $\tilde{E}_{b}=\mu_{b}$ since levels lower than {$\mu_b$} are already occupied,
whereas $\tilde{E}_{b}>\mu_{b}$ does not correspond to minimal depairing magnetic
field. To calculate $E_{f}$ we apply a variational approach maximizing
the probability of the potential $U(\mathbf{r})$ at a fixed value of the
fermion energy $E_{f}[ U( \mathbf{r})] $ in this
potential. The probability $P\left[ U\right] $ for the Gaussian
uncorrelated random potential is given by $P\left[ U\right] =\int DU\left(
\mathbf{x}\right) e^{-\frac{1}{2\kappa ^{2}}\int U^{2}\left( \mathbf{x}%
\right) d\mathbf{x}}.$ To find the optimal fluctuation (OF) for deeply
localized state, it is necessary to minimize $\int U^{2}\left( \mathbf{x}%
\right) d^{2}x$ at a fixed quantum particle energy $E\left[ \psi ,U\right]
=\int [ ({\hslash ^{2}}/{2m})\left( \nabla \psi \right) ^{2}+U\psi
^{2}] d\mathbf{x}$ \cite{Lifshitz}. The minimization over $U$ leads to
the relation $U=-\lambda \psi ^{2}$, where $\lambda $ is a Lagrangian
factor.  We choose the Gaussian fermion trial function $\psi _{f}(r)=(\alpha_f/{\pi}
)^{1/2}e^{-\alpha _{f}r^{2}/2\text{ }}$. All integrals can be calculated explicitly leading to the result: $\lambda =2\pi \hslash ^{2}/m$ and $\alpha
_{f}=-2mE_{f}/\hslash ^{2}$. Next we solve Schr\"{o}dinger equation for a boson in the potential $U_{b}({\bf x})
=2U_{f}=-2\lambda \psi _{f}^{2}$, using a similar
trial wave function and  require $\tilde{E}_{b}=\mu_{b}$.
This procedure gives $E_{f}=(2/9)\mu_{b}$.  The chemical potential of the
bosons $\mu_b$ was calculated in the works \cite{Falco2, babichenko}. For $d=2$ it reads
 $\mu_{b}\mathcal{=}-\EE_b\ln(\EE_b/gn_b)$ where $\EE_b=\hbar^2/(4m\LL_b^2)$. The SIT happens
at $n_bg\approx E_b$ where
{ $n_b\approx m\Delta /(8\pi\hbar^2)$}. Therefore, only the
disorder controls closeness to the SIT.  From  (\ref{BFT}), we arrive at the critical { parallel} field $B_{\textrm {BFT}\parallel}$ of a 2d film:%
\begin{equation}
\vspace{-0.cm}
B_{\textrm {BFT}}^{\Vert }\approx\frac{2\Delta -{5}\mu_{b}/9}{g_e\mu _{B}}={B_{c }%
}\Big[1+\frac{5}{18}\kappa%
\ln (\kappa/\kappa_c)
\Big].
\label{BFT-2par}
\end{equation}
\looseness-20 {H}ere $B_{c }={{2}\Delta }/({g_e\mu _{B}})$ is the CP breaking field.
$\kappa=\EE_b/\Delta$  measures the strength of the disorder relative to the gap; its value at the SIT is { $\kappa_c=gn_b/\Delta\approx mg/(8\pi\hbar^2)$}, which is the dimensionless (small) CP interaction constant.
\vspace{-0.cm}
\begin{figure}[h]
{\includegraphics[width=6cm]{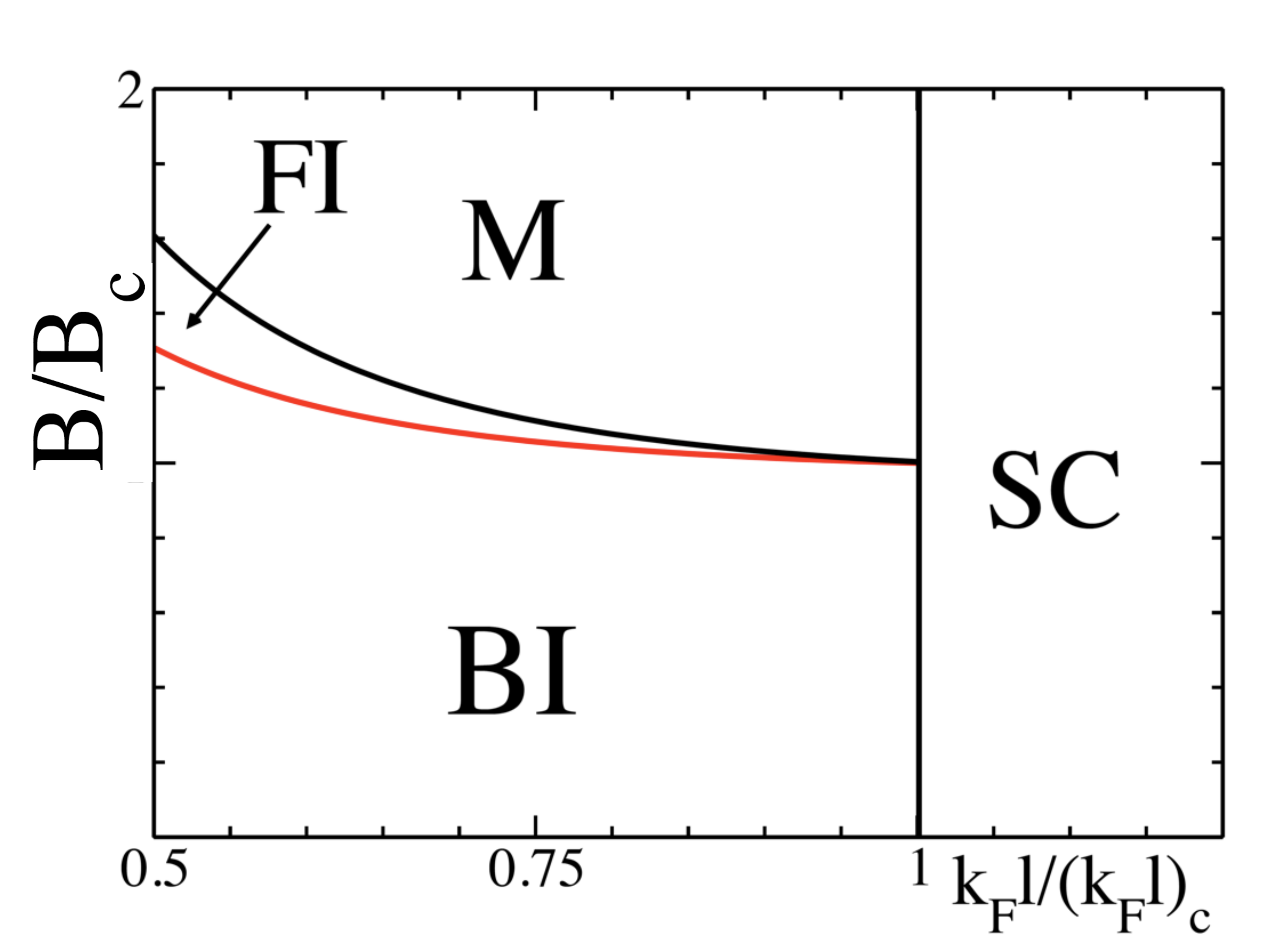}}
\vspace{-0.cm}
\caption{{\protect\small Phase diagram for disordered films in parallel
fields for $\gamma=1/2, (k_Fl)_c=1$ and $\kappa_c=0.2$. The vertical axis is the magnetic field in units of the CP breaking field, the horizontal axis $k_Fl/(k_Fl)_c$. BI, FI, SC and M  stand for
the Bose insulating, the Fermi insulating,  the superconducting  and the metallic phase, respectively. The red curve (online) corresponds to the 
 maximum of the resistivity. }}
\label{mit1}
\end{figure}
The bosons do not exist at
disorder so strong that the area of the optimal fluctuation $\pi\LL_b^2/\ln(\kappa/\kappa_c)$ \cite{Falco2} becomes less
than {$\pi\xi^2$} which  happens at
{$\kappa\approx 0.6/(k_Fl)$}.  For larger $\kappa$ (not shown in Fig. 1) there are no CPs in the insulating phase.  The squeezing line
together with the line (\ref{BFT-2par}) form the BFT line.
The fermion density in the FI phase is determined by eq. (\ref{BFT-2par}) in which the density of bosons must be taken $n_b=n-n_f/2$. At the MIT line $E_f=-\hbar^2n_{fc}/(2m)$ and $n_f=n_{ fc}=\LL_f{^{-2}}$. Thus
\begin{equation}
B_{\textrm {MIT}}^{\Vert }\approx B_{c }%
\left[1+\frac{\kappa}{ 2}\ln ({ \kappa}/({\kappa_c-\kappa/32}))\right].
\label{MIT-2par}
\end{equation}
The two lines of the BFT and MIT in the present approximation cross the SIT
line at the same point.
The MIT line goes to $B=\infty$ at $\kappa=32\kappa_c$. Since $\LL_b\sim l$ we rewrite $\kappa/\kappa_c\approx \sqrt{l_c/l}$ where
$(k_Fl)_c\approx 1$  at the SIT. A schematic phase diagram for the parallel field is shown in Fig. 1.

\noindent \looseness-30     \emph{Thin films in perpendicular field}.
 In this situation the
diamagnetic term in the Hamiltonian is nonzero. Still the wave functions
are isotropic and will be approximated as Gaussian
$\psi_k=\pi^{-1/2}\alpha_k^{1/2}\exp\left(-\alpha_kr^2/2\right)$. 
Otherwise the calculation follows the scheme discussed for the parallel field case.
In this way we find the
equation for the energy $E_k=-{\hbar^2\alpha_k}(1-3(2l_k^2\alpha_k)^{-2}))/2m_k$
and the density of potential wells $n_{w}\sim \alpha_k
 \exp[-\mathcal{L}_{k}^{2}\alpha_k (
1-(2\alpha \ell _{k}^{2}) ^{-2}) ^{2}]$
supporting the energy levels lower than a fixed $E_k$ \cite{Falco2}. 
 It interpolates between the two limiting cases of energy far below and very
close to the first Landau level \cite{Ioffe+81}.
To find the chemical potential of bosons one should minimize the total energy $\mu(\alpha_b)=E_b+gn_b/n_w(E_b)$
including the disorder and interaction contributions over $\alpha_b$ (see details in \cite{Falco2}).
For low fields, $B<
 \kappa B_c/(1-\gamma),$ the results for the case of parallel field apply, here $\gamma=1-m_0/mg_e$.
For larger fields  the extension of the  wave function is essentially
 determined by the magnetic length and the SIT line smoothly crosses over to the upper critical field $B_{c2}{
 = B_cd/(8k_Fl(1-\gamma))}=c_1\hbar c/(2e\xi^2)$, $c_1\approx0.69$.  At strong
magnetic fields, the density $n_{w}\left( E\right) $ rapidly {increases}
when the energy $E_{b}$ {approaches} the first Landau level.
\vspace{-0.4cm}\begin{figure}[h]
{\includegraphics[ width=6cm]{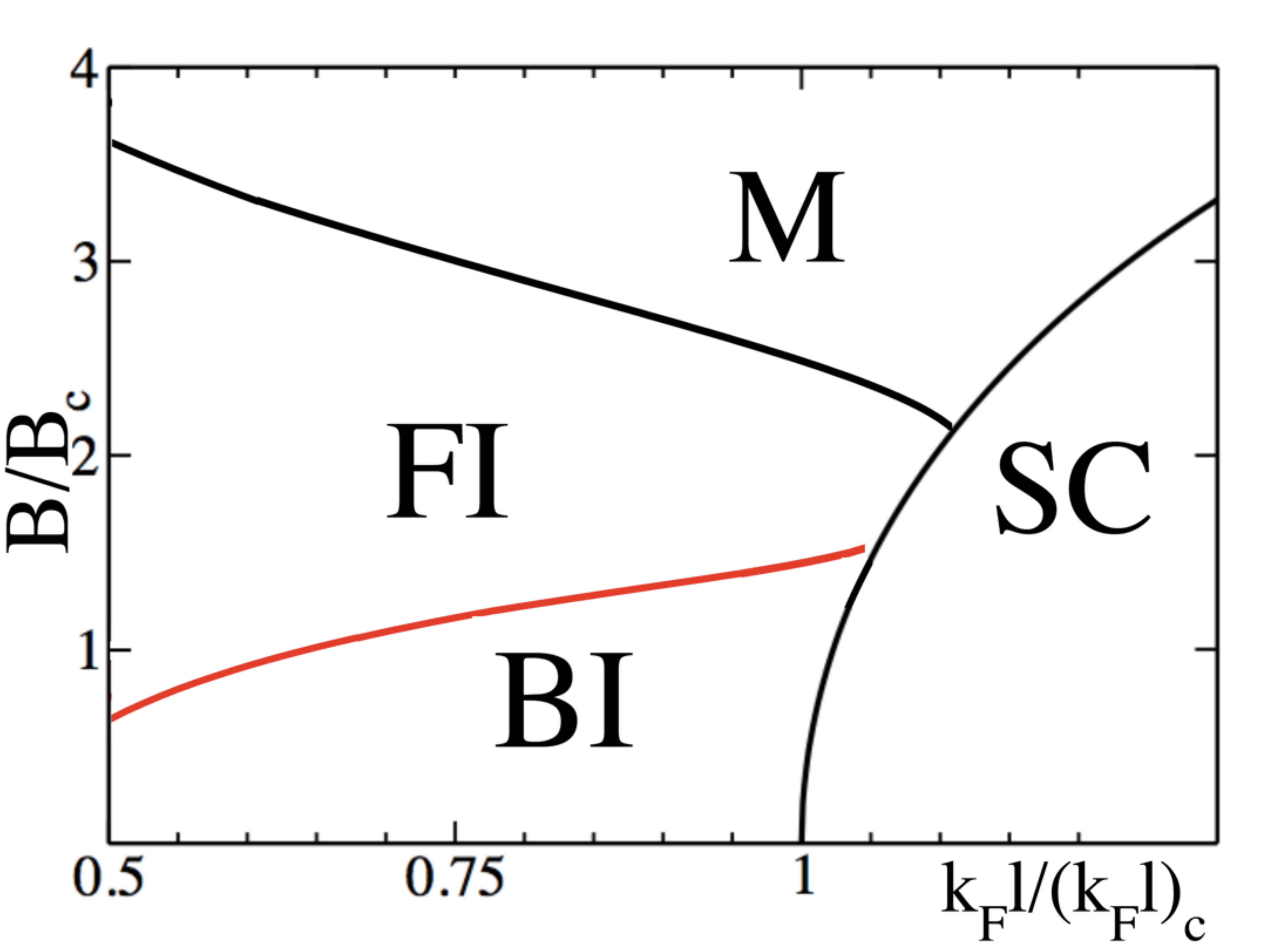}} 
\vspace{-0.2cm}
\caption{{\protect\small Phase diagram for disordered films in
 perpendicular field using the same parameters and notation as in Fig. 1. The BFT line is here due to squeezing.}}
\label{mit2}
\vspace{-0.2cm}
\end{figure}
The value $
E_{f}$ is calculated as before, but with the diamagnetic term. Plugging
these values into (\ref{BFT}), we
find the BFT and the MIT
lines for the strong perpendicular field.
Close to the SIT  the BFT line is  given  by $B_{\textrm{BFT}}^{\perp}\approx B_c[1+\sqrt{(\gamma^{-1}-1)\kappa/2\ln(\kappa/\kappa_c)}]/\gamma$. Far from the SIT line the result (\ref{BFT-2par}) applies with  the prefactor $5/18$  replaced by $5/9$.  The squeezing line $B_{sq}$ can be more relevant.
Its equation reads $\alpha _{b}=\xi^{-2}$.
This gives
 \begin{equation}
{B_{sq}^{\perp}\approx \frac{2{B_{c2}}}{c_1}
\left[1-\left(2c_1\kappa k_Fl\ln(\kappa/\kappa_c)
\right)^{\frac{1}{2}}\right]^{1/2}}
\end{equation}
At $\gamma{<}{2c_1}/({1+2c_1})$ and close to SIT ($k_Fl\approx 1$), the squeezing transition line passes below the paramagnetic one, but deeply in insulator region $k_Fl<1$ they interchange. It could explain a controversy  between the observed strong anisotropy of the resistance in magnetic field found in \cite{wu} and the absence of anisotropy found in \cite{adams-pairs}, both in Be films. The first experiment was done very close to the SIT, whereas the second one  deeply in insulator regime.
The phase diagram for this case is schematically depicted in
Figure 2.

\noindent\emph{Three-dimensional case. }\looseness=-20 In $d=3$ the
directions along and perpendicular to ${\bf B}$ are not equivalent.
Therefore, the trial wave functions are anisotropic Gaussian, $\psi
_{k}^{2}\left( \mathbf{x}\right) =\alpha _{k}\beta _{k}^{1/2}e^{-(\alpha _{k}%
\mathbf{\rho }^{2}+\beta _{k}z^{2})}/\pi ^{3/2}$.
The optimal
fluctuation has
ellipsoidal shape with long axis directed along magnetic field: $\beta
_{k}= \alpha _{k}-1/( 2\alpha _{k}\ell _{k}^{2})  $.
The squeezing transition proceeds when the longitudinal size of the
potential well $\beta _{k}^{-1/2}$ becomes equal to $\xi $. Otherwise the
calculations are analogous to those in the previous sections.
\vspace{-0.cm}
\begin{figure}[h]{\includegraphics[width=5cm]{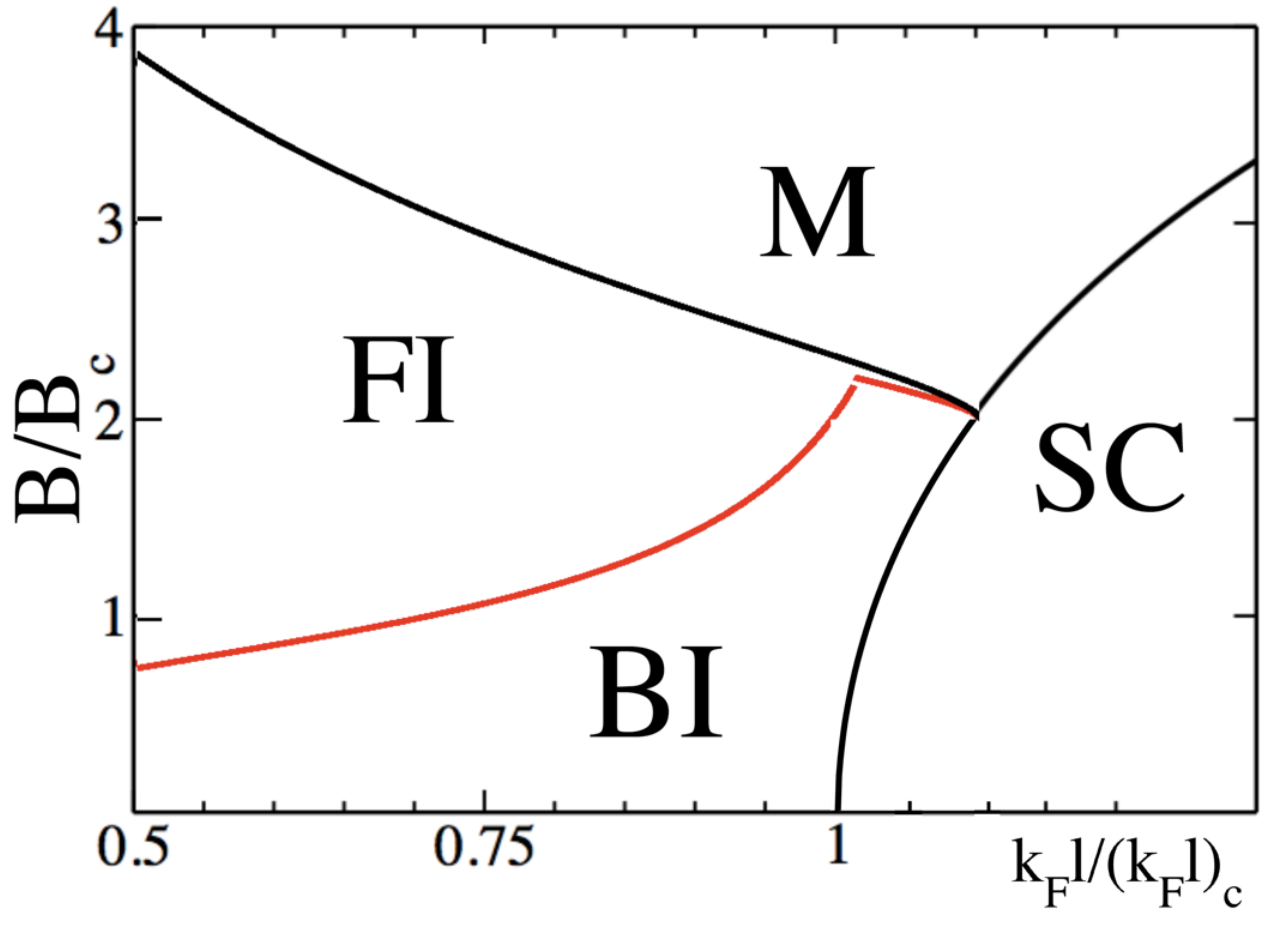}} %
{\includegraphics[width=4cm]{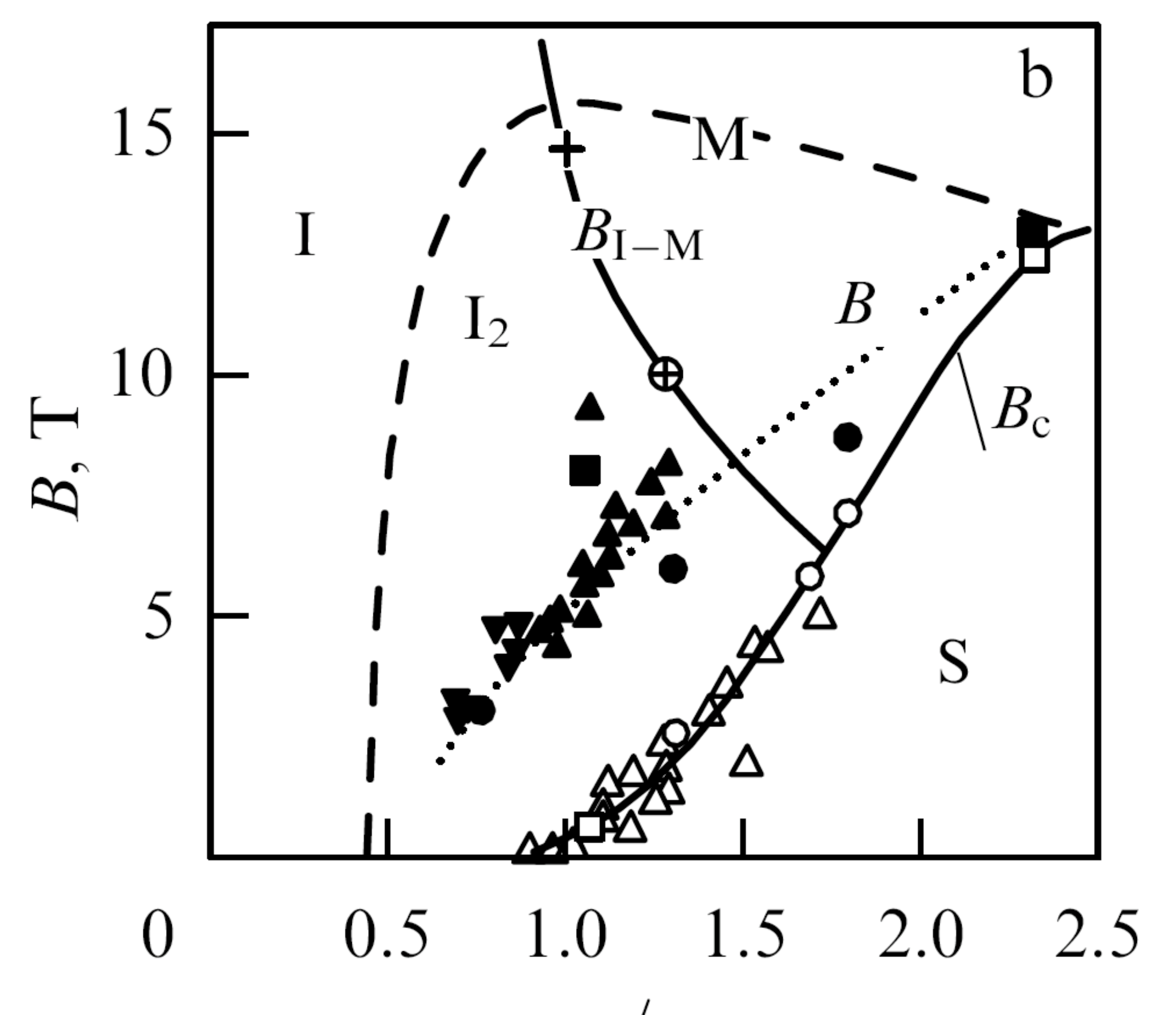}} 
\caption{{\protect\small Phase diagram for bulk samples for the parameters of Fig. 1. BI, FI, SC and M  stand for
the Bose insulating, the Fermi insulating,  the superconducting phase and the metallic phase, respectively. This diagram essentially reproduces topology of the experimental
phase diagram of \protect\cite{review} for In-O (right) .  The vertical axis of the latter is the magnetic field in Tesla, the horizontal axis is again $(k_Fl/(k_Fl)_c$.}}
\label{mit1}
\end{figure}The BFT line
consists of two parts. The squeezing transition {obeys equation:
\begin{equation}\label{3dsqueezing}
B_{\textrm{ sq}}^{3D}\approx\frac{ B_{c2}}{\sqrt{\kappa k_Fl}\ln (\kappa/\kappa_c)}\left[1-\sqrt{\kappa k_Fl}\ln (\kappa/\kappa_c)\right]^{1/2}
\end{equation}
 where $\kappa_c \sim mgk_F/\hbar^2$  is a small number. }
 At {decreasing  disorder}, the
squeezing curve crosses the paramagnetic BFT line whose equation for $\kappa\gg \kappa_c$ reads: $B_{\textsc{BFT}}^{3D}{\approx} B_{c}\left[1+0.2\kappa\ln^{2} (\kappa/\kappa_c)\right] $.  For larger densities
the BFT proceeds due to the Zeeman effect.
 At $\kappa\geq \kappa_{c}$ and  $\gamma<0$  the depairing is caused only by  squeezing, eq. (\ref{3dsqueezing}).
Equation of
the MIT line at {$\kappa\gg \kappa_{c}$} is given by
$B_{\textrm{MIT}}^{3D}\approx {B_{c}}\Big[ 1+{(\kappa/2)}\ln^2(\kappa/\kappa_c)\Big]$.
For { smaller disorder} one finds
\begin{equation}
B_{\textrm{MIT}}^{3D}=\frac{B_c}{\gamma}\left[ 1+\left(\frac{\kappa}{2}\ln^2{(\kappa/\kappa_c)}\right)^{1/3}\right].
\label{MIT-3}
\end{equation}

\noindent{\it Magnetoresistance.}  In zero magnetic field the wave functions decay on large scales exponentially, their overlap determines the hopping conductivity. For $B>0$ the decay of the wave function turns into  gaussian which leads to an increase of the resisteivity \cite{Shklovskii}. This behavior proceeds until fermions appear in the FI phase where the 
resistance sharply decreases in comparison to the BI phase. Indeed, {if} both
resistances have variable range hopping nature, {they obey} the Mott
formula: $R=R_{0}\exp [-\left( T_{0}/T\right) ^{1/\left( d+1\right) }]$. The
difference between fermions and bosons is in the value of $T_{0}\mathcal{
\simeq }\alpha ^{d/2}\mathcal{E}/n_{w}\left( \mu \right) \sim \mathcal{E}
/\left( n\mathcal{L}^{d}\right) $. The ratio of the critical temperatures
reads $T_{0f}/T_{0b}\sim ({n_{b}}/{n_{f}})\left( {\mathcal{L}_{b}}/{\mathcal{
L}_{f}}\right) ^{d+2}$. At $n_{f}\sim n_{cf}=\mathcal{L}_{f}^{-2}$ one finds
$T_{0f}/T_{0b}\sim \left( n_{b}/n_{cb}\right) \left( \mathcal{L}_{b}/
\mathcal{L}_{f}\right) ^{2}.$ It is small even in close vicinity of the SIT.
In InO$_{x}$ experimenters observed the activation behavior of resistance at
low fields and the VRH behavior at high fields $\sim 10-15$T. The natural
explanation is that the CPs have an intrinsic energy gap modified by
disorder, whereas there is no gap for electrons.

\noindent \emph{ Conclusions.}To conclude, we demonstrated that both the BFT and MIT transitions should
happen in 2d, though the phase diagrams are very different for parallel and
perpendicular field. We predict a strong anisotropy with respect to the
direction of magnetic field in 2d close to the SIT which must be eventually
suppressed by increasing disorder. 3d systems (films 20nm and thicker) are
isotropic with respect to the magnetic field direction. They always display
the BFT transition, but the MIT transition does not happen if $\gamma<0$.
Our phase diagram in 3d has the same topology as the experimental phase
diagram established in the review \cite{review} by Gantmakher and Dolgopolov.

\noindent

This work has been supported by SFB 608 (T.N. and V.P.) and by the DOE under
the grant DE-FG02-06ER 46278 (V.P.). {We are indebted to P. Adams, T. Baturina
A. Finkelstein, M. Feigelman, and V. Gantmakher for discussions and B. Roostaei for his help in creating the  figures.}


\begin{thebibliography}{99}
\bibitem{shahar-92} D. Shahar and Z. Ovadyahu, Phys. Rev. B \textbf{46, }
10917 (1992).

\bibitem{hebard} A.F. Hebard and M.A. Paalanen, Phys. Rev. Lett. \textbf{65},
927 (1990); M.A. Paalanen, A.F. Hebard, and R.R Ruel, Phys. Rev. Lett.
\textbf{69}, 1604 (1992).

\bibitem{gant-2000} V. F. Gantmakher et al.,
JETP Lett. \textbf{68}, 363 (1998), ibid. \textbf{71}, 160 (2000).
\bibitem{gant-01} V.F. Gantmakher and M.V. Golubkov, JETP Lett.  \textbf{73}, 131 (2001).
%
\bibitem{baturina} T.I. Baturina et al., JETP Lett. \textbf{79}, 337 (2004), T.I. Baturina et al. Phys. Rev. Lett. \textbf{99}, 257003 (2007).
%
\bibitem{beschoten} B. Beschoten et al.,
Phys. Rev. Lett. \textbf{77}, 1837 (1996).
%
\bibitem{goldman} D.B. Haviland, Y. Liu, and A.M. Goldman, Phys. Rev. Lett.
\textbf{62}, 2180 (1989).
%
\bibitem{adams} P.W. Adams, P. Herron, and E.I. Meletis, Phys. Rev. B
\textbf{58}, R2952 (1998).
%
\bibitem{butko} V.Yu. Butko and P.W. Adams, Nature \textbf{409}, 161 (2001).
%
\bibitem{wu} {}E. Bielejec, J. Ruan, and Wenhao Wu, Phys. Rev. B \textbf{63},
100502 (2001); E. Bielejec and Wenhao Wu, Phys. Rev. Lett. \textbf{88},
206802 (2002).
%
\bibitem{valles} J.M. Valles et al., Phys. Rev. B \textbf{39}, 11599 (1989);
Y. Ando et al., Phys. Rev. Lett. \textbf{87}, 017001 (2001).
%
\bibitem{review} V.F. Gantmakher and V.T. Dolgopolov, Physics-Uspekhi
\textbf{53}, 1 (2010).
%
\bibitem{shahar-04} G. Sambandamurthy et al.,
Phys. Rev. Lett. 92, 107005 (2004).
%
\bibitem{Gantmakher-96} V.F. Gantmakher, M.V. Golubkov, J.G.S. Lok, and A.K.
Geim, Sov. Phys. JETP \textbf{82}, 951 (1996).
%
\bibitem{valles-09} H.Q. Nguyen \textit{et al., }Phys. Rev. Lett. \textbf{103%
}, 157001 (2009).
%
\bibitem{adams-pairs} Y.M. Xiong, A.B. Karki, D.P. Young, and P.W. Adams,
Phys. Rev. B \textbf{79}, 020510 (2009).
%
\bibitem{kapitulnik} M.A. Steiner, N.P. Breznay, and A. Kapitulnik, Phys.
Rev. B \textbf{77}, 212501 (2008).
%
\bibitem{finkelstein} A. M. Finkelstein, JETP Letters \textbf{45}, 46 (1987).
%
\bibitem{efetov} K. B. Efetov, Sov. Phys. JETP \textbf{51}, 1015 (1980).
%
\bibitem{Shklovskii-09} M. M\"{u}ller and B.I. Shklovskii, Phys. Rev. B
\textbf{79}, 134504 (2009).
%
\bibitem{fisher} M.P.A. Fisher, Phys. Rev. Lett. \textbf{65}, 923 (1990).
%
\bibitem{ghosal} A. Ghosal, M. Randeria, and N. Trivedi, Phys. Rev. B
\textbf{65}, 014501 (2001).
%
%
\bibitem{meir2} Y. Dubi, Y. Meir, and Y. Avishai, Nature \textbf{449}, 876
(2007); Phys. Rev. B \textbf{78}, 024502 (2008).
%
\bibitem{priority} The existence of FI and BI was first proposed in the work
%
\cite{hebard} on the basis of observation of two different MIT transitions
for the dissipative and Hall resistances, which to our knowledge was not
confirmed by other experiments.
%
\bibitem{Falco2} G.M. Falco, T. Nattermann, and V.L. Pokrovsky, Europhys.
Lett. \textbf{85}, 30002 (2009); Phys. Rev. B. {\textbf{80}}, 104515 (2009).
%
\bibitem{LM} M. Ma and P.A. Lee, Phys. Rev. B \textbf{32}, 5658 (1985);
L.N. Bulaevsky and M.A. Sadovsky, Pis'ma ZhETF \textbf{39}, 524 (1984);
A. Kapitulnik and G. Kotliar, Phys. Rev. Lett. \textbf{54}, 473 (1985).
%
\bibitem{babichenko} A. Babichenko and V. Babichenko, Phys. Lett. A {\textbf{%
373}}, 2973 (2009).
%
\bibitem{Lifshitz} I. M. Lifshitz,
Sov. Phys. JETP \textbf{26}, 462 (1968).
%
\bibitem{Ioffe+81} L.B. Ioffe and A.I. Larkin, Sov. Phys. JETP \textbf{54}, 556 (1981).
{Our numerical factors in exponent are  about $1.2$ times larger than exact results.}

\bibitem{Shklovskii} B.I. Shklovskii and A.L. Efros, {\it Electronic Properties of Doped Semiconductors},
Springer, Berlin (1984).


\end{thebibliography}
\end{document}